\documentclass[aps,prl,twocolumn,superscriptaddress,nofootinbib,amsmath,amssymb,floatfix]{revtex4}
\pdfoutput=1
\usepackage{graphicx}  
\usepackage{dcolumn}   
\usepackage{bm,color}        
\usepackage{amssymb}   

\newcommand{\be}{\begin{equation}}
\newcommand{\ee}{\end{equation}}

\newcommand{\rd}{\mathrm{d}}
\newcommand{\rmd}{\mathrm{d}}

\usepackage{soul}
\usepackage{enumitem}

\definecolor{myBlue}{RGB}{15, 40, 220}
\usepackage[linktocpage=true]{hyperref}
\hypersetup{colorlinks=true, citecolor=myBlue, linkcolor=myBlue, urlcolor=violet}

\begin{document}

\title{Implications of the scalar tilt for the tensor-to-scalar ratio}

\author{Paolo Creminelli}
\affiliation{Abdus Salam International Centre for Theoretical Physics,\\ Strada Costiera 11, 34151, Trieste, Italy}
\author{Sergei Dubovsky}
\affiliation{Abdus Salam International Centre for Theoretical Physics,\\ Strada Costiera 11, 34151, Trieste, Italy}
\affiliation{Center for Cosmology and Particle Physics, Department of Physics, New York University New York, NY, 10003, USA}
\author{Diana L\'opez Nacir}
\affiliation{Abdus Salam International Centre for Theoretical Physics,\\ Strada Costiera 11, 34151, Trieste, Italy}
\affiliation{Departamento de F\'isica and IFIBA, FCEyN UBA, Facultad de Ciencias Exactas y Naturales, Ciudad Universitaria, Pabell\'on I, 1428 Buenos Aires, Argentina}

\author{Marko Simonovi\'c}
\affiliation{Institute for Advanced Study, Princeton, NJ 08540, USA}
\author{Gabriele Trevisan}
\affiliation{SISSA, via Bonomea 265, 34136, Trieste, Italy}
\affiliation{Istituto Nazionale di Fisica Nucleare, Sezione di Trieste, 34136, Trieste, Italy}
\author{Giovanni Villadoro}
\affiliation{Abdus Salam International Centre for Theoretical Physics,\\ Strada Costiera 11, 34151, Trieste, Italy}
\author{Matias Zaldarriaga}
\affiliation{Institute for Advanced Study, Princeton, NJ 08540, USA}


\begin{abstract}
We investigate the possible implications of the measured value of the scalar tilt $n_s$ for the tensor-to-scalar ratio $r$ in slow-roll, single-field inflationary models. The measured value of the tilt satisfies $n_s -1\sim 1/N_*$, where $N_* \sim 60$ is the number of $e$-folds for observationally relevant scales. If this is not a coincidence and the scaling holds for different values of $N$, it strongly suggests that either $r$ is as big as $10^{-1}$ (a possibility in tension with the recent data), or smaller than $10^{-2}$ and exponentially dependent on $n_s$. A large region of the ($n_s$,$r$) plane is not compatible with this scaling.
\end{abstract}

\maketitle

{\em Introduction.---}Planck confirmed previous indications that the spectrum of scalar perturbations is not scale invariant: $n_s-1 = -0.032 \pm 0.004 \;$ at $1\,\sigma$ \cite{Planck:2015xua}. This is surely an important step in the understanding of the early Universe:  inflation generically predicts a deviation from scale invariance, although the magnitude is, as we will discuss, model dependent. The experimental value of $|n_s-1|$ is of order $1/N_* \simeq 0.017$, where $N_*$ is the number of $e$-folds to the end of inflation for observationally relevant scales (we are going to take $N_* = 60$ for definiteness). This did \emph{not} have to be the case: it is easy to find models on the market with $|n_s-1|$ much bigger, say $0.2$ (of course the slow-roll approximation requires the tilt to be much smaller than $1$), or much smaller, say $10^{-4}$. For example in the prototypical hybrid inflation model
\be
V = \frac12 m^2\phi^2 + \frac14 \lambda (\psi^2- M^2)^2 + \lambda' \phi^2\psi^2
\ee
the tilt is $n_s-1 \simeq  2\eta = (2 m^2 M_{\rm P}^2)/V_0$, where $V_0 = \frac14 \lambda M^4$ is the vacuum energy during inflation, before the field $\psi$ relaxes to the true minimum. The tilt is a constant and does not depend on $N$: it can be much smaller or much larger than $1/N$. (In this example the tilt is positive, but the same applies to inverted hybrid models with red tilt.) In this kind of models, the inflaton ``does not know" when inflation is going to end, i.e.~when the waterfall field will become tachyonic. Thus there is no relation between the tilt, which only depends on the derivatives of the potential at a given point, and $N$, which measures the distance to the end of inflation. The approximate equality $n_s-1 \sim 1/N$ could just be  an accident.

On the other hand in this note we want to take this indication seriously and see what are the implications on inflation, and in particular on the expected amount of gravitational waves. Our formulas will be similar to Refs.~\cite{Roest:2013fha} and \cite{Mukhanov:2013tua} (see also Refs.~\cite{Boyanovsky:2005pw} and \cite{Boubekeur:2014xva}) although the implications we will draw will be slightly different.
\vskip.3cm
{\em Main argument.---} The experimental value of the scalar tilt suggests
\be
\label{basic}
n_s -1 = - \frac{\alpha}{N}\;,
\ee
with $\alpha$ of order unity. We assume the equation above to be valid in a window which is comfortably larger than the observable one: in other words the same equation would hold if one were to measure perturbations at, say, $N = 10$ or $N =200$ instead of $N=60$. For the time being we assume $\alpha$ is strictly a constant and later discuss deviations from this assumption. Writing the tilt in terms of $\epsilon \equiv - \dot H/H^2$ and its derivative, the equation above becomes (at first order in slow roll) a differential equation for $\epsilon$
\be
\label{diffe}
n_s -1 = - 2 \epsilon + \frac{\rmd \log\epsilon}{\rmd N} = - \frac{\alpha}{N} \;.
\ee
This is easily integrated to give
\be
\label{epsilon}
\epsilon(N) = \frac1{2(\alpha-1)^{-1} N + A N^\alpha} \;,
\ee
with $A$ an integration constant which can be related to the number of $e$-folds $N_\times$ where the two terms in the denominator are equal: $A=2 N_\times^{1-\alpha}/(\alpha-1)$.  By a judicious choice of $A$ (or equivalently $N_\times$) one can choose any value for $\epsilon$ (and thus for $r$) at $N_*=60$. However the scaling \eqref{basic} 
says that there is nothing special at the scale $N_*=60$ we measure, and therefore it looks reasonable to further assume that, in a certain parametric window around $N_*$, only one of the two power laws at the denominator of Eq. \eqref{epsilon} dominates. Conversely $N_*=60$ would  be accidentally close to the transition point between the two regimes. 

 Within this assumption one has two different cases depending on whether $\alpha$ is larger or smaller than 1. For $\alpha > 1$ there are two possible behaviors, depending on which scaling of $\epsilon$ is chosen, while only  one solution exists for $\alpha < 1$,  since $\epsilon$ cannot be negative. 
Therefore there are three cases:
\vskip.2cm
1) $\alpha>1$ and $\epsilon \simeq (\alpha-1)/2N$.  The value of $\epsilon$ (and thus of $r$) is fixed and large. This is the case of monomial potentials $V \propto \phi^{2\alpha-2}$. This is the simplest and most informative scenario: inflation is driven by a simple monomial potential, $r$ is large enough to make $\%$ measurements possible \cite{Dodelson:2014exa} (see also Refs.~\cite{Lee:2014cya,Creminelli:2015oda}) and we would be quite confident of what is going on \cite{Creminelli:2014oaa, Creminelli:2014fca}. However, the most recent bounds on $r$ \cite{Array:2015xqh} almost exclude this scenario at $2\sigma$.

\begin{figure}[t!]
\centering
\includegraphics[width=\columnwidth]{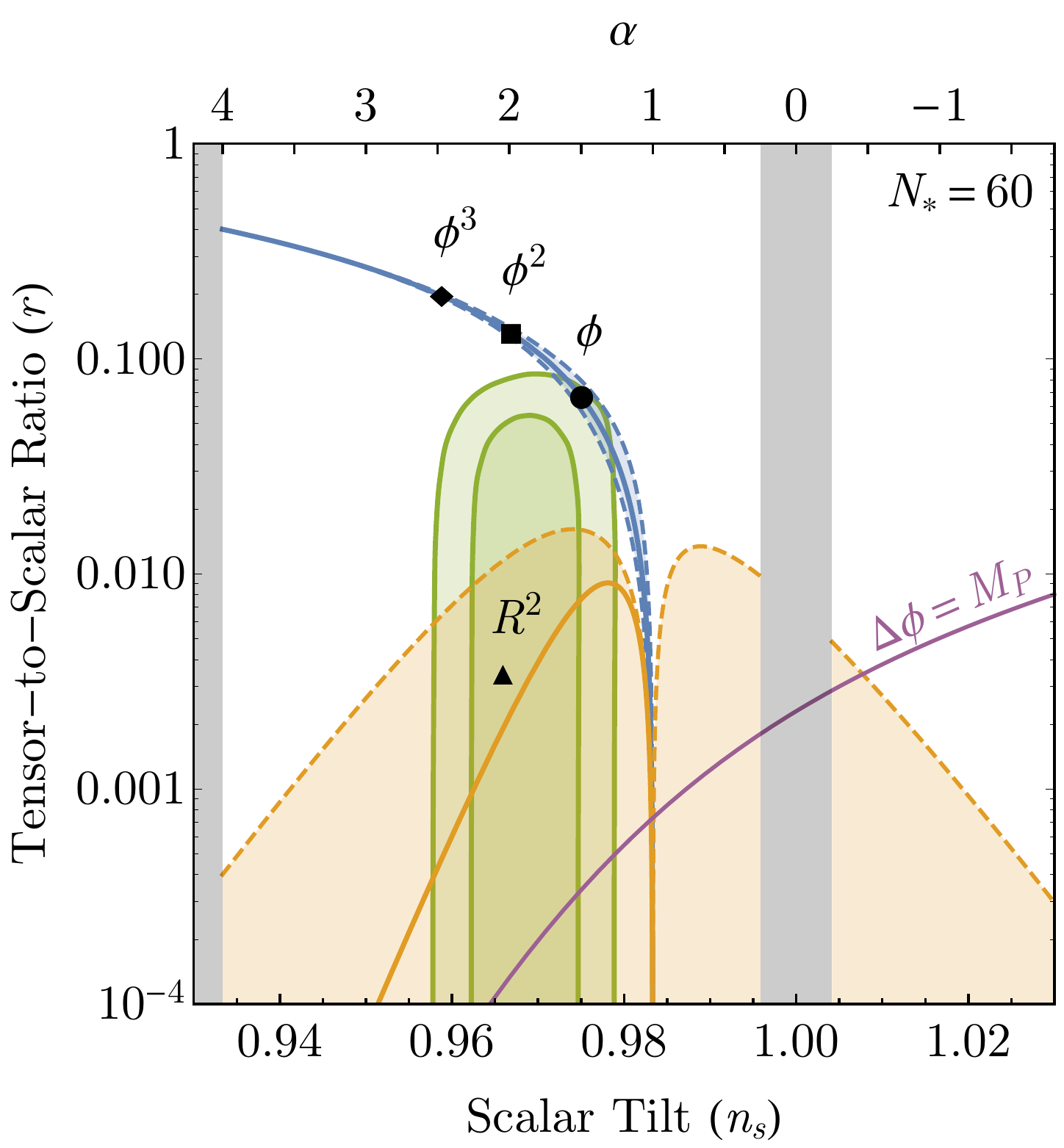}
\caption{\it \small{Possible allowed regions in the experimental ($n_s$,$r$) plane, as derived from our assumptions. The solid blue and orange lines correspond to the behavior of case 1) and 2). Dashed lines depend on the choice of $\bar N$ and $N_\times$, as explained in the text. The experimentally allowed region is in green (1 and 2\,$\sigma$ contours). In the gray shaded regions $|\alpha|$ is significantly different from one, so that the assumption in Eq.~(\ref{basic}) may not apply. The solid purple line corresponds to the Lyth bound.}}
\label{fig:bounds}
\end{figure} 

\vskip.2cm
2) $\alpha>1$ and $\epsilon \simeq  A^{-1} N^{-\alpha}$. In this case one cannot fix the value of $\epsilon$: the only requirement is that the constant $A$ is big enough so that one can neglect the first term at the denominator of \eqref{epsilon}. In terms of $r$ this gives
\be
\label{smallr}
r = 16 \epsilon\simeq16 A^{-1} N^{-\alpha} \lesssim 8(\alpha-1) N_\times^{\alpha-1}N^{-\alpha} \;.
\ee
Of course, $r$ depends not only on $\alpha$ but also on $N_\times$. As already stated, we are assuming that the crossing is far away from our observable window, e.g. $N_\times\simeq1$ (solid orange line in Fig.~\ref{fig:bounds}), or $N_\times\simeq N_*/10$ (left dashed orange line in Fig.~\ref{fig:bounds}).

It is easy to find the potentials that correspond to these behaviors \cite{Garcia-Bellido:2014gna}.

The case $\alpha >2$ consists for example of hilltop models that inflate around the origin,
\be
V(\phi) = V_0 \left[1- \left(\frac{\phi}{M}\right)^n\right]\;,
\ee
with $n>2$ and $M \lesssim M_{\rm P}$. For these potentials $\alpha = (2n-2)/(n-2)$. Notice that for $n = 2$ the potential does not follow the $1/N$ scaling, since $\eta$ (and thus $n_s$) goes to a constant at small $\phi$.\footnote{This suggests that if one modifies the potential with $n = 2$ with a correction which goes to zero slower than any polynomial, one gets intermediate behavior for $n_s-1$. For example the potential
$V = V_0[1+(\phi/M)^2 /\,{\rm{log}} (\phi/M)]$ gives $n_s-1\propto1/\sqrt{N}$. This shows that the $1/N$ scaling is not the only possibility, although arguably the most natural. It is the experimental value of $n_s$ that suggests $1/N$.}

For $\alpha = 2$ one has models that approach a constant exponentially for large $\phi$
\be
\label{phimn}
V(\phi) = V_0 \left[1- e^{-\phi/M}\right] \;,
\ee 
with $M \lesssim M_{\rm P}$.\footnote{Notice that the potentials we are quoting for each case are just examples and that there are completely different potentials giving the same $\alpha$. For instance the potential $V_0 [1-\exp{(M/\phi)}]$ near the origin gives the correct $1/N$ scaling with $\alpha =2$.}

In the case $1< \alpha < 2$ one finds models that approach a constant polynomially at large $\phi$ 
\be\label{phim-n}
V(\phi) = V_0 \left[1- \left(\frac{M}{\phi}\right)^n\right] \;,
\ee 
with $n>0$ and $M \lesssim M_{\rm P}$. For these $\alpha = 2(n+1)/(n+2)$. 

The potentials given above are just examples which reproduce approximately Eq.~\eqref{basic}. For example in the case above of models that approach a constant polynomially, corrections to Eq.~\eqref{basic} go as $(N^{-1} \cdot M^2/M_{\rm P}^2)^{n/(2+n)}$: for $M \lesssim M_{\rm P}$, this is a good approximation to Eq.~\eqref{basic} (unless $n$ is too small, see later). 

In all the cases $A^{-1} \sim (M/M_{\rm P})^{2\alpha-2}$. As we discussed, this number cannot be large, see Eq.~\eqref{smallr}, but unfortunately it can be arbitrarily small, when the scale $M$ is smaller than the Planck scale. As done in Refs.~\cite{Mukhanov:2013tua,Roest:2013fha} one can assume that $M \simeq M_{\rm P}$, or equivalently that $\epsilon \simeq N^{-\alpha}$, however this is an additional assumption and not a consequence of Eq.~\eqref{basic}. For smaller values of $M$ (and thus of $\epsilon$) slow roll terminates because $\eta$ becomes of order one: after that $\epsilon$ starts varying fast and reaches unity in one $e$-fold or so. For example brane inflation corresponds in its simplest form to a potential of the form of Eq.~\eqref{phim-n} with $n=4$ and $M$ parametrically smaller than $M_{\rm P}$ \cite{Baumann:2014nda}. Furthermore, exponential potentials are ubiquitous in field and string theory constructions, both with $M \sim M_{\rm P}$ and with $M \ll M_{\rm P}$ \cite{Baumann:2014nda,Burgess:2013sla}.
\vskip.2cm
3) $\alpha<1$ and $\epsilon \simeq  A^{-1} N^{-\alpha}$. This regime is qualitatively different from the previous ones. The second term in the denominator of Eq.~\eqref{epsilon} must dominate, since the first term would give a negative $\epsilon$. Since the first term grows faster than the second for large $N$, this case cannot be sustained for arbitrarily large $N$. On the other hand we can require that it is valid for a large window around the observable scales, say up to $\bar N = 10N_*$. Again this gives an upper bound on the amplitude of gravitational waves
\be
r = 16 \epsilon \lesssim \frac{1-\alpha}{2} \frac{1}{\bar N} \left(\frac{\bar N}{N}\right)^\alpha \;.
\ee
Although it may look artificial, this behavior can be obtained with the potential
\be
V = V_0 \left[1+ \left(\frac{\phi}{M}\right)^n\right]\;,
\ee
with $0 < n < 2$ and $M\gg M_{\rm P}$. Equation~\eqref{basic} is valid with  $\alpha = -2(n-1)/(2-n)$ for $\phi \ll M$ (as we said this regime cannot last for arbitrarily large $N$).

This class covers also the case of blue tilt of order $1/N$.  But there is an important difference: for red tilt slow roll breaks when $\eta$ becomes large and negative, so that $\epsilon$ is increasing and naturally leads to the end of inflation $\epsilon =1$. For blue tilt $\eta$ is large and positive at the end of slow roll, so that $\epsilon$ is small and decreasing. Some additional ingredient is needed to ultimately terminate inflation. Of course this is not so interesting since a blue tilt is ruled out experimentally. 
Notice also that one should not use our arguments too close to the scale-invariant point $|\alpha| \ll 1$ because this would violate the assumption that $n_s-1$ is of order $1/N$. The same applies when $|\alpha|$ becomes parametrically larger than one. Anyway, both these cases are experimentally ruled out. 

In Fig.~\ref{fig:bounds} we draw the various possibilities together with the current experimental bounds \cite{Planck:2015xua} and the predictions for power-law potentials and the Starobinsky model \cite{Starobinsky:1980te}, requiring that one of the two behaviors of $\epsilon$ dominates in a window up to $\bar N = 10N_*$. The solid orange line corresponding to Eq.~\eqref{smallr} is defined only up to a factor of order unity. The dashed line for $\alpha <1$ and around the large-$\epsilon$ solution depends on $\bar N$, and should be thus interpreted with care. We also draw (solid purple line) the Lyth bound \cite{Lyth:1996im}: in Eq.~(\ref{epsilon}) we choose the value of $A$ such that the displacement of the inflaton from $N_*=60$ to the end is $\Delta\phi=M_{\rm P}$ \cite{Garcia-Bellido:2014wfa}. Within the experimentally allowed region for the scalar tilt, all ``measurable" values of $r$ ($\gtrsim5\times10^{-4}$) correspond to $\Delta\phi>M_{\rm P}$.

It is important to stress that the $1/N$ scaling, given in Eq.~\eqref{basic}, can be checked experimentally by the measurement of the running which is obviously fixed to be $\alpha_s = -\alpha/N_*^2 \simeq - 7 \times 10^{-4}$ \cite{Garcia-Bellido:2014gna}. Unfortunately this value is probably too small to be measured with 
of a larger running would disprove Eq. (2).
cosmic microwave background (CMB) experiments \cite{Andre:2013nfa}; however a measurement of a larger running would disprove Eq.~\eqref{basic}.

\vskip.3cm
{\em Stability of the constraints.---}Of course one cannot argue from the measurement of the tilt that Eq.~\eqref{basic} holds with $\alpha$ strictly constant. At most one can argue that $\alpha(N)$ is a slowly varying function of $N$ (\footnote{It is easy to find examples of potentials where there are corrections to the exact $1/N$ scaling: for example the potential $V =V_0[1-{\rm exp}\,(-\phi^2/M_{\rm P}^2)]$ has $n_s-1\simeq-2/N\cdot(1+1/(2\,{\rm log}\,N))$. }). Let us check that the qualitative features of the plot in Fig.~\ref{fig:bounds} remain the same. If $\alpha$ depends on $N$, Eq.~\eqref{diffe} can be written as a linear differential equation 
\be
\frac{\rd \epsilon^{-1}}{\rd \log N} - \alpha(N) \epsilon^{-1} =  - 2 N \;,
\ee
whose general solution is
\begin{equation}
\begin{split}
\epsilon^{-1}(N) = &-2 e^{\int_1^N \!  \frac{d \tilde N}{\tilde N} \alpha(\tilde N)} \int_1^N \!\!d \tilde N e^{-\int_1^{\tilde N} \! \frac{d \hat N}{\hat N} \alpha(\hat N)} \\ &
+ A e^{\int_1^N \! \frac{d \tilde N}{\tilde N} \alpha(\tilde N)} \;.
\end{split}
\end{equation}
The first line on the rhs is the non homogeneous solution and it reduces to $2(\alpha-1)N$ for constant $\alpha$. When $\alpha$ is not a constant the solutions will  not be power laws, but we can still assume that one of the two behaviors (corresponding to $\epsilon \sim \eta$ or $\epsilon \ll \eta$) dominates over a parametric window without moving from one to the other. The constraints on $\epsilon$ (and thus on $r$) will be perturbatively close to the case of constant $\alpha$ if the variation is small. 
The second line corresponds to the homogeneous solution of the differential equation. It amounts to neglecting the  contribution of $\epsilon$ to the tilt, $\epsilon \ll \eta$, and it reduces to the power-law $N^{\alpha}$ for a constant $\alpha$. 

If $\alpha$ weakly depends on $N$, the plot of Fig.~\ref{fig:bounds} will be slightly ``blurred". For example if Eq.~\eqref{basic} is modified to allow for a ``running'' $\alpha$
\be
n_s -1 = - \frac{\alpha}{N}\left(\frac{N}{N_*}\right)^\delta\;,
\ee
then the power law solution $\epsilon\propto N^{-\alpha}$ is modified for small $\delta$ by a factor $(1+ \alpha \delta (\log N_*)^2/2 + \ldots)$. If we take $\delta \simeq 0.3$, in such a way that the effective $\alpha$ changes by a factor of 2 as $N$ varies by an order of magnitude, the correction is of order $2$. This does not affect our conclusions, since Eq.~\eqref{smallr} is anyway defined up to a factor of order unity. These uncertainties will sum up with the experimental uncertainties on $n_s$ and the theoretical ones on the number of $e$-folds $N$.  This in particular tells us we should not take too seriously the small value of $r$ in the region close to $\alpha =1$: the two solutions $N$ and $N^\alpha$ becomes closer and closer and the results are very sensitive to the corrections we just discussed.
\vskip.3cm
{\em Conclusions.---}The robust conclusion is that there are regions in the  $(n_s, r)$ plane which are {\em not} compatible with the $1/N$ hypothesis of Eq.~\eqref{basic} (see also Ref.~\cite{Roest:2013fha}), and the assumption that there is no change of behavior for $\epsilon$. Unfortunately these assumptions do not set a lower bound for $r$. If one further requires that $\epsilon$  becomes of order one when slow roll breaks, then we have either the case 1) or the case 2) with the inequality \eqref{smallr} saturated (solid orange line in Fig.~\ref{fig:bounds}). Conversely the $1/N$ scaling is compatible with an arbitrarily low energy during inflation. In particular it is also compatible with large-$f_a$ QCD axion models, which would be in tension with high-scale inflation models \cite{Fox:2004kb, Seckel:1985tj,Beltran:2006sq}.

It is important to stress that, since in Eq.~\eqref{smallr} $r$ depends exponentially on the tilt, an improvement on the experimental limits of this quantity will be of great importance.

Current and upcoming CMB experiments will be able to probe values of $r$ as small $2\times 10^{-3}$ \cite{Creminelli:2015oda} in the not-so-distant future. If experiments will put us in the ``forbidden" region, we will have to give up one of the assumptions. One possibility is that the value of the tilt is only accidentally of order $1/N$. Inflation requires the slow-roll parameters to be small, but in explicit constructions it may be difficult to make them as small as we like. For example supergravity corrections (or in general Planck-suppressed operators) tend to push $\eta$ towards one ($\eta$ problem), thus giving large contributions to the tilt. Similarly it appears difficult to have pseudo-Nambu-Goldstone bosons with a decay constant much bigger than $M_{\rm P}$ \cite{ArkaniHamed:2006dz}, so that a not-so-small tilt of order $M_{\rm P}^2/f^2$ is expected. One can surely reproduce the tilt we observe in these cases, though one might argue that a larger value would be expected if the flatness of the potential is so hard to maintain. Another way out is a small speed of sound for the inflaton. Current constraints still allow a substantial reduction in the value of $r$. The other assumption we might have to relax is that $\epsilon$ does not move from one behavior to the other close to our observable window. For example in Ref.~\cite{Kallosh:2013yoa} the authors considered the model $V \propto \tanh^{2}(\phi/\sqrt{6\beta})$ which satisfies, for any $\beta$, Eq.~\eqref{basic} with $\alpha =2$. One can obtain values of $r$ in the forbidden region by adjusting $\beta$ in such a way that the two terms in the denominator of Eq.~\eqref{epsilon} are comparable for $N = N_*$. However, this requires some amount of tuning since observable inflation happens very close to the inflection point of the potential.
\vskip.3cm
\emph{Acknowledgements.---}We would like to thank D. Roest for useful discussions. S.D. is supported in part by the NSF CAREER award PHY-1352119, M.S. gratefully acknowledges support from the Institute for Advanced Study, the work of G.V. is partly supported by the ERC Advanced Grant no. 267985 ``DaMESyFla'', and M.Z. is supported in part by NSF Grants No. PHY-1213563 and No. AST-1409709.




\begin{thebibliography}{99}  

\bibitem{Planck:2015xua} 
  [Planck Collaboration],
  arXiv:1502.01589 [astro-ph.CO].
  
  

\bibitem{Roest:2013fha} 
  D.~Roest,
  JCAP {\bf 01}, 007 (2014)
  [arXiv:1309.1285 [hep-th]].

\bibitem{Mukhanov:2013tua} 
  V.~Mukhanov,
  Eur.\ Phys.\ J.\ C {\bf 73}, 2486 (2013)
  [arXiv:1303.3925 [astro-ph.CO]].
  
\bibitem{Boyanovsky:2005pw} 
  D.~Boyanovsky, H.~J.~de Vega and N.~G.~Sanchez,
  Phys.\ Rev.\ D {\bf 73}, 023008 (2006)
  [astro-ph/0507595].
  
\bibitem{Boubekeur:2014xva} 
  L.~Boubekeur, E.~Giusarma, O.~Mena and H.~Ramírez,
  arXiv:1411.7237 [astro-ph.CO].
  
\bibitem{Dodelson:2014exa} 
  S.~Dodelson,
  Phys.\ Rev.\ Lett.\  {\bf 112}, 191301 (2014)
  [arXiv:1403.6310 [astro-ph.CO]].
  
\bibitem{Lee:2014cya} 
  H.~Lee, S.-C.~Su and D.~Baumann,
  arXiv:1408.6709 [astro-ph.CO].
  
\bibitem{Creminelli:2015oda} 
  P.~Creminelli, D.~L.~L\'opez Nacir, M.~Simonovi\'c, G.~Trevisan and M.~Zaldarriaga,
  JCAP {\bf 1511}, no. 11, 031 (2015)
  doi:10.1088/1475-7516/2015/11/031
  [arXiv:1502.01983 [astro-ph.CO]].

\bibitem{Creminelli:2014oaa} 
  P.~Creminelli, D.~L\'opez Nacir, M.~Simonovi\'c, G.~Trevisan and M.~Zaldarriaga,
  Phys.\ Rev.\ Lett.\  {\bf 112}, 241303 (2014)
  [arXiv:1404.1065 [astro-ph.CO]].
  
\bibitem{Creminelli:2014fca} 
  P.~Creminelli, D.~L\'opez Nacir, M.~Simonovi\'c, G.~Trevisan and M.~Zaldarriaga,
  Phys.\ Rev.\ D {\bf 90}, 083513 (2014)
  [arXiv:1405.6264 [astro-ph.CO]].
  
\bibitem{Array:2015xqh} 
  P.~A.~R.~Ade {\it et al.} [BICEP2 and Keck Array Collaborations],
  arXiv:1510.09217 [astro-ph.CO].
  
\bibitem{Garcia-Bellido:2014gna} 
  J.~Garcia-Bellido and D.~Roest,
  arXiv:1402.2059 [astro-ph.CO].
  

  

  

  
\bibitem{Baumann:2014nda} 
  D.~Baumann and L.~McAllister,
  arXiv:1404.2601 [hep-th].
  
\bibitem{Burgess:2013sla} 
  C.~P.~Burgess, M.~Cicoli and F.~Quevedo,
  JCAP {\bf 1311}, 003 (2013)
  [arXiv:1306.3512 [hep-th]].
  
\bibitem{Starobinsky:1980te} 
  A.~A.~Starobinsky,
  Phys.\ Lett.\ B {\bf 91}, 99 (1980).
  
\bibitem{Lyth:1996im} 
  D.~H.~Lyth,
  Phys.\ Rev.\ Lett.\  {\bf 78}, 1861 (1997)
  [hep-ph/9606387].
  
\bibitem{Garcia-Bellido:2014wfa} 
  J.~Garcia-Bellido, D.~Roest, M.~Scalisi and I.~Zavala,
  arXiv:1408.6839 [hep-th].
 
\bibitem{Andre:2013nfa} 
  P.~Andr {\it et al.}  [PRISM Collaboration],
  JCAP {\bf 1402}, 006 (2014)
  [arXiv:1310.1554 [astro-ph.CO]].
 
\bibitem{Seckel:1985tj} 
  D.~Seckel and M.~S.~Turner,
  Phys.\ Rev.\ D {\bf 32}, 3178 (1985).
  
\bibitem{Fox:2004kb} 
  P.~Fox, A.~Pierce and S.~D.~Thomas,
  hep-th/0409059.
  
\bibitem{Beltran:2006sq} 
  M.~Beltran, J.~Garcia-Bellido and J.~Lesgourgues,
  Phys.\ Rev.\ D {\bf 75}, 103507 (2007)
  doi:10.1103/PhysRevD.75.103507
  [hep-ph/0606107].
  
\bibitem{ArkaniHamed:2006dz} 
  N.~Arkani-Hamed, L.~Motl, A.~Nicolis and C.~Vafa,
  JHEP {\bf 0706}, 060 (2007)
  [hep-th/0601001].
  
\bibitem{Kallosh:2013yoa} 
  R.~Kallosh, A.~Linde and D.~Roest,
  JHEP {\bf 1311}, 198 (2013)
  [arXiv:1311.0472 [hep-th]].
  
  
 


  

  
      

\end{thebibliography}
\end{document}